\documentclass[%
 reprint,
nofootinbib,
 amsmath,amssymb,
 aps,
]{revtex4-2}

\usepackage[utf8]{inputenc}
\usepackage{hyperref}
\usepackage{amsmath}
\usepackage{color}
\usepackage{subfig}
\usepackage{comment}
\usepackage{wrapfig}
\usepackage{graphicx}
\usepackage{tabularx}
\usepackage{booktabs}
\usepackage{multirow}

\begin{document}
\preprint{APS/123-QED}

\title{Point Spread Function Deconvolution Using a Convolutional Autoencoder}

\author{Sreevarsha Sreejith$^{1,3}$}
\author{An\v{z}e Slosar$^{1}$}
\author{Hong Wang$^{2}$}

\affiliation{$^{1}$ Physics Department, Brookhaven National Laboratory, Upton, NY 11973}
\affiliation{$^{2}$ Computational Science Initiative, Brookhaven National Laboratory, Upton, NY 11973}
\affiliation{$^{3}$ Physics Department, University of Surrey, Guildford GU2 7XH, UK}

\begin{abstract}

A major issue in optical astronomical image analysis is the combined effect of the instrument's point spread function (PSF) and the atmospheric seeing that blurs images and changes their shape in a way that is band and time-of-observation dependent. In this work we present a very simple neural network based approach to non-blind image deconvolution that relies on feeding a Convolutional Autoencoder (CAE) input images that have been preprocessed by convolution with the corresponding PSF and its regularized inverse, a method which is both conceptually simple and computationally less intensive. We also present here, a new approach for dealing with limited input dynamic range of neural networks compared to the dynamic range present in astronomical images.

\end{abstract}
\maketitle

\section{Introduction} \label{sec:intro}
Astronomy is entering into an era of big data with existing and upcoming ground-based and space-borne observatories probing remarkably large volumes of the observable sky with depths and cadences that are hitherto unseen. With upcoming surveys such as the Legacy Survey of Space and Time (LSST, \cite{lsstbook2009}) to be conducted at the Vera C. Rubin observatory, combined with existing data sets from the Dark Energy Survey (DES, \cite{des2018}) \& Hyper-Suprime Cam (HSC, \cite{hsc2019}), and space-based surveys to be conducted by James Webb (JWST, \cite{jwst2006}), Euclid (\cite{euclid2010}) \& Nancy Grace Roman (\cite{roman2012}) telescopes, this `big data overload' is only expected to become more overwhelming. Both space-borne and ground-based telescopes have certain hurdles to overcome in terms of resolution. While space-based observatories are diffraction limited, ground-based telescopes have to cope with atmospheric seeing. Atmospheric seeing occurs when an optical wavefront passes through atmospheric turbulence and  the perturbations in the wavefront causes the image to get distorted, resulting in a finite point spread function (PSF). Since the field of view is large, it is not yet feasible to implement adaptive optics across the focal plane, leading to a PSF that is arcsecond-sized.

Since the atmosphere surrounding the telescope is variable, the PSF is expected to vary in size, shape and orientation from one observation to another. It also varies according to the passbands used for imaging and is generally asymmetric. Fortunately, PSF is precisely known for each exposure because stars contained in each exposure are excellent point sources and can be used to directly measure the PSF.

 In the face of varying PSF and noise, we want to recover unbiased shapes, positions and fluxes of astronomical objects and machine learning techniques have been used in recent years to great effect for this. Most of these methods have been focused on galaxy deblending. \cite{reiman2019deblending} proposed a branched deblender that makes use of generative adversarial networks (GANs) to deblend overlapping field galaxies. \cite{boucaud2020photometry} explored the usage of a simple convolutional neural network (CNN) in conjunction with a U-Net (\cite{unet2015}) in order to do image segmentation/galaxy deblending and measure photometry. \cite{arcelin2021} developed a variational autoencoder (VAE) - like network and we ourselves explored using residual dense neural networks (RDN) for the same purpose (\cite{wang2021}). 

The ideal end product for such machine-learning approaches would be to homogenize the dataset to the extent that would make the subsequent data analysis considerably simpler. Realistic astronomical observations are done in a variety of observing conditions that lead to variable depth and PSF size and shape. These get imprinted into any quantities that are measured directly from the images. The most canonical example is galaxy weak-lensing, where major efforts have been spent trying to understand how the PSF shape and size and image noise levels affects the measurement \cite{2018ARA&A..56..393M}.  One could imagine a fundamental change in the approach where the images are deconvolved and denoised using a neural-network approach and the subsequent quantities are derived using simpler shear estimation algorithms on processed images. The beauty of this approach is that it would allow many more morphological quantities to be used in a robust manner to improve cosmological analysis, either by using marked correlation functions \cite{2006MNRAS.369...68S} (where we are marking on some morophological quantity of interest) or to perform non-canonical analysis such as correlation of the apparent galaxy spin directions (see e.g. \cite{2009MNRAS.392.1225S}). Of course, the current state of the art in terms of neural-network based approach is still well behind the state of the art of traditional astronomical image analysis. In this paper we consider one of the many aspects that need to be resolved before such approach can become reality. 

Most of the existing approaches are formulated with the assumption of a constant PSF, with the caveat that the model could be retrained/modified with techniques like transfer learning with different data sets with varying PSFs. This could be impractical for the reasons that we have laid out in our earlier paper, \cite{wang2023} (\texttt{HW23} henceforth). Moreover, the assumption of a constant PSF could lead to biased photometry for individual objects. Therefore, in order to apply neural network approaches to real world scenarios, it is necessary to address the issue of how to treat image PSFs (in addition to other problems such as artefacts, blending, masking etc.). In this work, as in \texttt{HW23}, we focus specifically on tackling the PSF issue and propose a simple neural network architecture that, with minimal image preprocessing, removes the residual PSF dependence and recovers object positions and shapes effectively.
 
\section{Synthetic Data} \label{sec:data}

The base data set that is used in this work was simulated in the same manner as in \texttt{HW23}, using the deep generative models proposed in \cite{lanusse2021}. These models are generated using a combination of a hybrid variational autoencoder (\cite{kingmawelling2013}) with the aggregate posterior distribution modelled by a latent-space normalizing flow, termed as \texttt{Flow-VAE}, and was trained on a data set based on the HST/ACS COSMOS survey (\cite{Koekemoer_2007,Scoville_2007a}) and rendered using GalSim (\cite{galsim2015}). The physical parameters that are to be specified for image generation are half-light radius (\texttt{flux\_radius}, as a proxy for size), apparent magnitude in the i-band (\texttt{mag\_auto}, derived from SExtractor, \cite{ba1996}) and photometric redshift (\texttt{zphot}). The values for these parameters were drawn from the respective uniform distributions as, $5 \leq \texttt{flux\_radius} \leq 15$, $5 \leq \texttt{mag\_auto} \leq 25$ and $0 \leq \texttt{zphot} \leq 2$. In order to make the galaxy placement in the images  more realistic, we slightly offset the objects from the centre in both \texttt{x} and \texttt{y} directions with a value randomly drawn from a uniform distribution between \texttt{(-5,5)}. 

The first major difference from the data set used in \texttt{HW23} is that we do not convolve the generated galaxy image with a small non-functional PSF that aliases and removes the modes that might lead to over-deconvolution. Since the deep generative model only provides a single band (band 1) and we are interested in multi band deconvolutions, we generate two more bands with simple non-linear transformations of band 1 (see Equation 24 and the $\alpha$ and $\beta$ values  in \texttt{HW23}) and pile them along the third axis to form a 3-band image, which makes up the `truth' data set for this work. Then each individual band of the truth data set is convolved with a random Moffat PSF ($0.6 \leq \texttt{FWHM} \leq 1$ arcsecond, $2 \leq \beta \leq 5$, $-0.9 \leq g1,g2 \leq 0.9$, $\sigma_{g1,g2} = 0.1$, $\mu_{g1,g2} = 0$), making the PSF different both across the three bands for one object and also across the entire data set. Then random Gaussian noise ($\texttt{mean} = 0$, $0 \leq \texttt{variance} \leq 0.1$, $\sigma = \texttt{variance} ^ {0.25}$ ) was added to create the noisy images. This is the second major difference from \texttt{HW23} --instead of using the \texttt{addNoiseSNR} function from \texttt{Galsim}, we manually add noise to the images. The pixel scale of the PSF of band 1 is set as the pixel scale of the 3-band image (varies in the range 0.088 - 0.255 pixels/arcseconds over the data set). The signal-to-noise ratio (SNR) of the data set is calculated as Equation 25 in \texttt{HW23} and it ranges between $1.6$ and $130$. 
 
In addition to the exclusion of the small Gaussian PSF, the difference in the noise addition and the difference in the ranges of the SNR, in this data set we also add a few `blank' images (30,000 in all -- 27,000 for training \& 3000 for testing) to the truth data set in contrast to the \texttt{HW23} data set. Addition of these blanks will help us determine how the network would perform in conditions of very low SNR, or when spuriosities in noisy images might appear object-like even though there is no actual object. All-in-all our data set consists of $117,000$ objects in the training set and $13,000$ objects in the test set including `blanks' and `non-blanks'. 

\section{Approach} \label{sec:method}

Convolutional auto-encoders (CAE) are an attractive method for analyzing astronomical images because they are inherently translationally invariant. One can train them to perform a deconvolution for a fixed PSF, where the shape of that PSF gets burned into weights that do the deconvolution. But which approach should one take if the PSF shape is free and is one of the  inputs to the problem?

Simply feeding the PSF image in addition to the actual noisy input image will not do the job in an ordinary convolutional network, since they respect locality: features around a certain coordinate $(x,y)$ are combined in some complicated non-linear, but still approximately local manner to produce the output image, while the relation between the PSF and the noisy image is convolutional:

\begin{equation}
    I_{i} = I_{gt,i} \circledast PSF_i + N_i, \label{eq:problem}
\end{equation}
where $I_i$ denotes the observed image, $I_{gt}$ the ground truth and $N_i$ the noise image.
However, we can let operations that appear in classical image analysis guide our intuition. These are,

\newcommand{\xPSF}{\texttt{$\times$PSF}}
\newcommand{\xiPSF}{\texttt{$\times$iPSF}}
\newcommand{\xiPSFO}{\texttt{$\times$iPSF1}}
\newcommand{\blur}{\texttt{blur}}

\begin{figure*}[!ht]
    \centering
    \includegraphics[width=\linewidth]{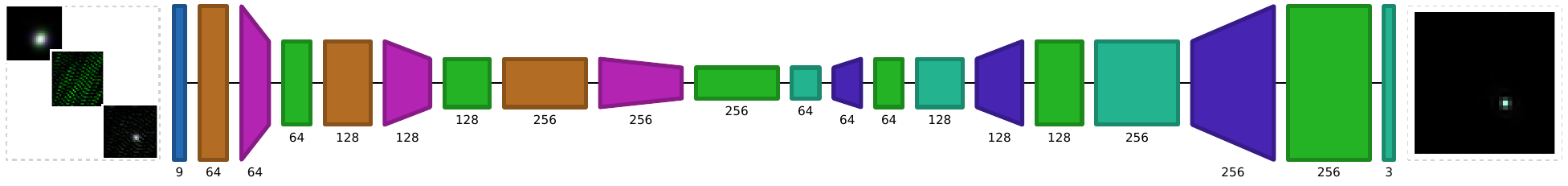} \\[15pt]
    \includegraphics[width=\linewidth]{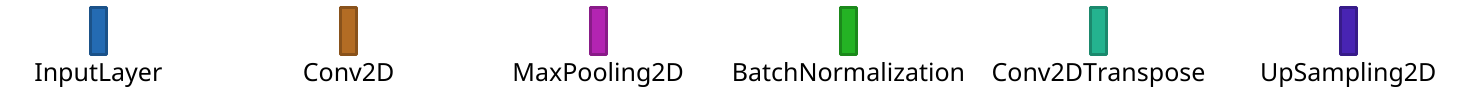} \\
    \caption{Figure showing the schematic representation of the CAE  used in this work. The input image is a compound data cube of \xPSF, \xiPSF and \xiPSFO with dimensions $32\times 32\times 9$ per image. The output image dimensions are $32\times 32\times 3$. The numbers shown indicate the change in dimensions of the image tensor as it propagates through the network. The different components of the network are indicated in the legend. The network diagram was generated using \cite{net2vis}.}    
    \vspace{1cm}
    \label{fig:nwd}
\end{figure*}

\begin{enumerate}
    \item \textbf{Convolution with PSF}, which we denote as \xPSF. Convolution of the noisy image with PSF acts as a matched filter for object detection and is optimal for point sources and approximate  for small sources. However, it makes any shape distortion due to assymetric PSF even worse.
\item \textbf{Convolution with inverse PSF}, which we denote as \xiPSF. Convolution with the inverse PSF (formally defined as a function which, when convolved with PSF produces a delta function response) is the naive solution to the deconvolution problem. When not regularized, it is expected to amplify noise, but since the standard regularization is to convolve again with a regularizing kernel (typically a Gaussian), we anticipate that the network would learn this process by itself. 
\item \textbf{Convolution with the inverse PSF convolved with a 1 arcsec Gaussian PSF}, which we denote as \xiPSFO. This is a regularized version of the above, which can make it better, but also looses information. The size of the regularizing kernel was chosen after several trial and error scenarios, from which it was concluded that $1$ is the standard deviation at which the network performs the best. We have confirmed that our results are stable across reasonable range of regularizing kernel sizes.  
\end{enumerate}

For each actual input galaxy we produced these 3 types of manipulated data. Each image of each of these types has a size of $32 \times 32 \times 3$ (trimmed down from the $35 \times 35 \times 3$ used in \texttt{HW23} and preprocessed in the same manner with maximum value scaling).  While in \texttt{HW23} the maximum values across each object for the noisy and truth data sets was used for scaling, in our case the maximum values across each object of the \xPSF, \xiPSF, \xiPSFO and the truth data is used. In this manner, the minimum and maximum values of all four data sets fall between 0 and 1. We experimented with various combinations of those inputs to identify those that work optimally and therefore our network input size varied from $32\times 32\times 3$ to  $32\times 32\times 9$.

\subsection{Convolutional Autoencoder}
The network used in this work is a simple 2-dimensional convolutional autoencoder (CAE) with three layers ($\times 2$ combinedly for the encoder and the decoder parts). 

Autoencoders (AE, \cite{ae1985}) are a ``self-supervised'' machine learning method that is generally used in image processing for dimensionality reduction and feature extraction with the encoder and decoder parts being separate neural networks with similar configurations. The encoder takes the input data and maps it to a latent space (encoded space) which is generally of a lower dimensionality than the input data, thus achieving data compression and dimensionality reduction. The decoder decompresses/recompresses the data, often with some associated loss which is dealt with by neural networks in the normal fashion, such as by using backpropagation to update weights between batches.  The bottleneck thus constructed allows for the clearly structured portion of the data to pass through, which brings about effective denoising (an important consideration, especially in astronomical images). In the case of a single layer AE, this framework is analogous with principal component analysis (PCA, \cite{pearson1901,hotelling1933}) provided it uses a linear activation function. When the AE consists of multiple layers, thus making it deep and non-linear, the loss during the decoder phase becomes more complex. This is the advantage AE has over PCA, it's ability to learn non-linear patterns in the input data with lower dimensions and less data loss. 
 As mentioned earlier, in this work we have implemented a CAE, which is simply an AE with convolutional layers  and a convolutional bottleneck, and in our case, is trained end-to-end\footnote{There are also AEs that are trained layer by layer, but these are a different type of AEs called ``stacked'' AEs.}.  More details about the specific network architecture is given in the following section.
 
\subsection{Network architecture}
Our implementation of the CAE is based on the Keras library in Python \footnote{https://keras.io/} within Tensorflow framework \footnote{https://www.tensorflow.org/}, utilising the available APIs. The encoder layer consists of three convolutional layers and three pooling layers with the convolutional layers activated with a `Leaky ReLU' (\cite{maas2013leaky}, leaky rectified linear unit) function with $\alpha = 0.1$.  In normal ReLU \citep{relu2011}, the negative part of the function is set to 0 which renders the unit inactive. This causes what is called `dying ReLU' problem that can sometimes lead to overfitting.  In Leaky ReLU this is dealt with by applying a non-zero slope, indicated by $\alpha$. All $6$ layers have padding set to `same' and the pooling utility used for spatial downsampling is `Maxpooling'. The kernels used are $5\times5$ in the convolutional layers and $2\times2$ in the pooling layers. The convolutional layers have respective filter sizes of 64, 128 and 256. The decoder layers are symmetric to the encoder layers except that a transpose convolution is applied and that instead of the pooling layers there is an `UpSampling' layer per transpose convolution layer for spatial upsampling, which also utilises a $2 \times 2$ kernel.  The final output layer has 3 units with a $3\times3$ kernel, `same' padding and activated with a `softplus' function \citep{nair2010spre}. Figure \ref{fig:nwd} shows a schematic representation of the network.

The regularization aspect of the network is achieved by using `BatchNormalization' layers in both the encoder and decoder portions. There is much discussion in literature as to whether dropout is necessary for CNNs or whether batch normalization is useful for autoencoders. In this case, based on trial and error we have decided to simply apply  `BatchNormalization'  layers after each pooling/upsampling layer as might be the case in the final network. The model thus formulated is compiled with the \texttt{Adam} optimizer \citep{adam2014} and the loss function \texttt{BinaryCrossentropy} (BCE)\footnote{https://www.tensorflow.org/api\_docs/python/tf/keras/losses/BinaryCrossentropy}, both from the Keras API. BCE is generally used in image classification problems rather than reconstruction, we opted for it in this work because of its superior performance to the others that we tested (see Table \ref{bcemse} in the Appendix and the explanation thereof). The initial learning rate is $0.0001$ and it decays exponentially at the rate of $0.6$ per $100000$ steps. The model is trained for $350$ epochs with a batch size of $10$.
We intend for this method to be a more specific methodological sequel to the work in \texttt{HW23} in order to provide unbiased shear measurements that do not correlate with the PSF but do so with the truth image. This simple, `quick and dirty' data-driven approach  provides excellent results, which are illustrated here using the same metrics those were used in \texttt{HW23} in the following sections. It is also cheaper in terms of computation time and resources, taking $\sim 3 $ hours to train on an NVIDIA GeForce GTX 1650 GPU with 4GB memory and compiled with cuDNN\footnote{https://developer.nvidia.com/cudnn} (version 11.4) . 
\begin{figure*}[htbp]
\begin{minipage}{\linewidth}
\centering
\scalebox{0.95}{
\begin{tabular}{c}
\includegraphics[width=\linewidth]{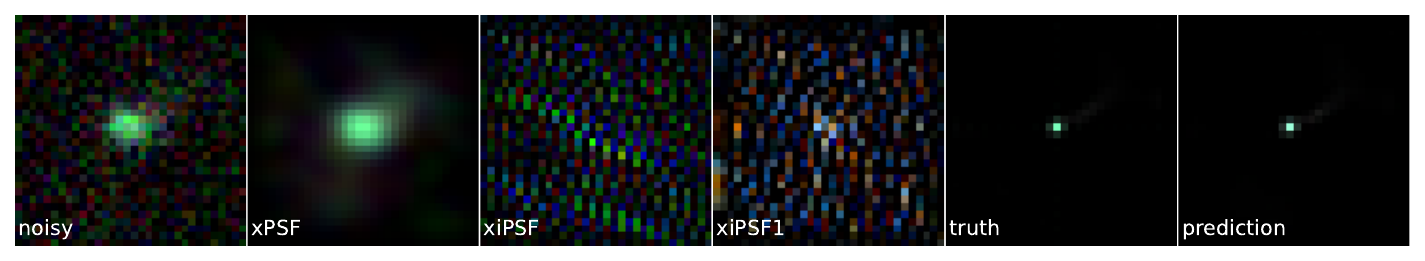}\\
\includegraphics[width=\linewidth]{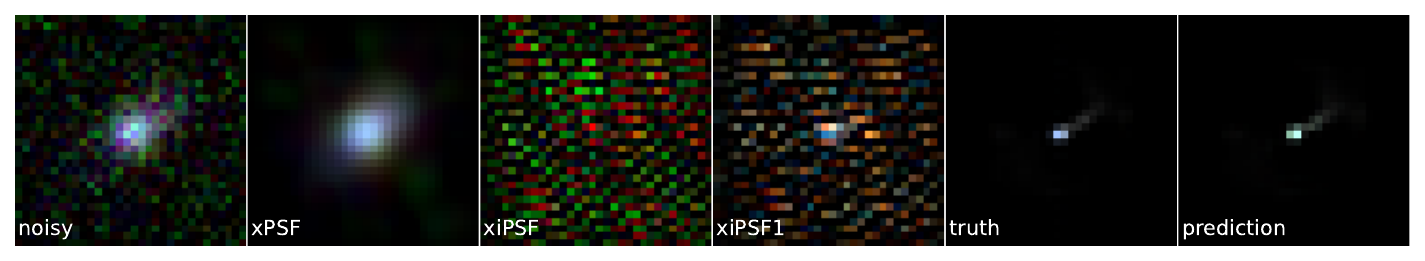}\\
\end{tabular}
}\\
\caption*{Low SNR}
\end{minipage}
\\
\begin{minipage}{\linewidth}
\centering
\scalebox{0.95}{
\begin{tabular}{c}
\includegraphics[width=\linewidth]{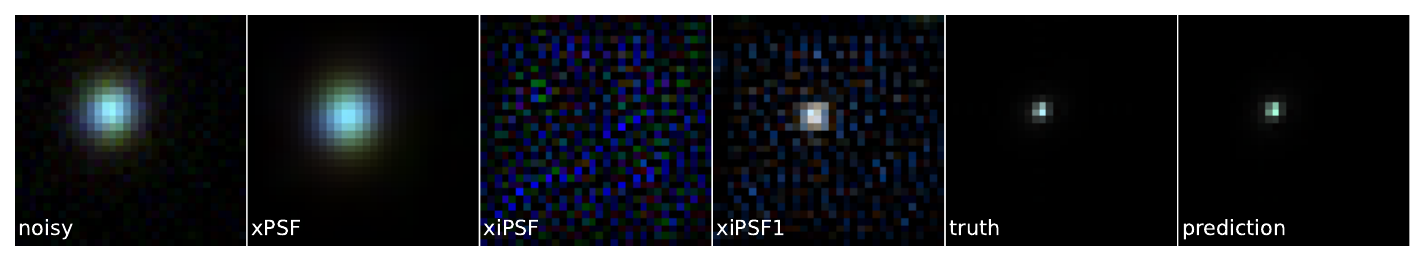}\\
\includegraphics[width=\linewidth]{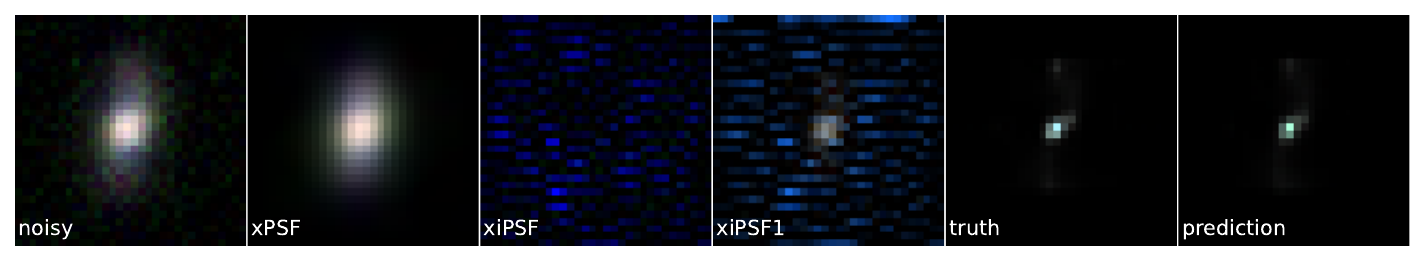}\\
\end{tabular}
}\\
\caption*{Medium SNR}
\end{minipage}
\\
\begin{minipage}{\linewidth}
\centering
\scalebox{0.95}{
\begin{tabular}{c}
\includegraphics[width=\linewidth]{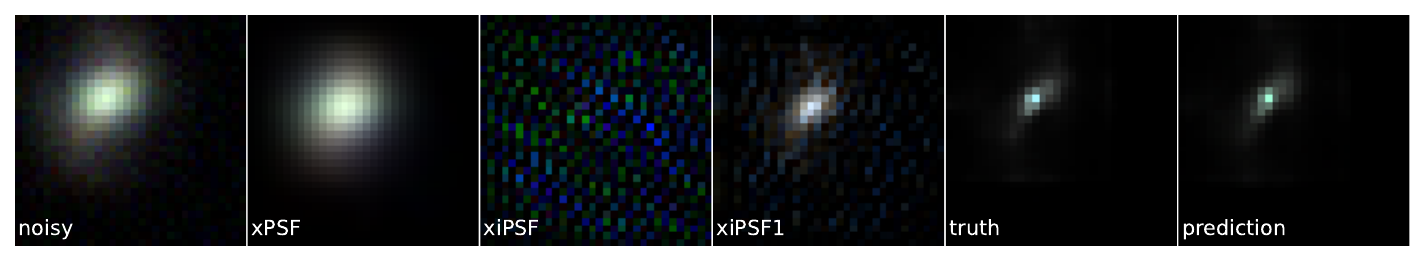}\\
\includegraphics[width=\linewidth]{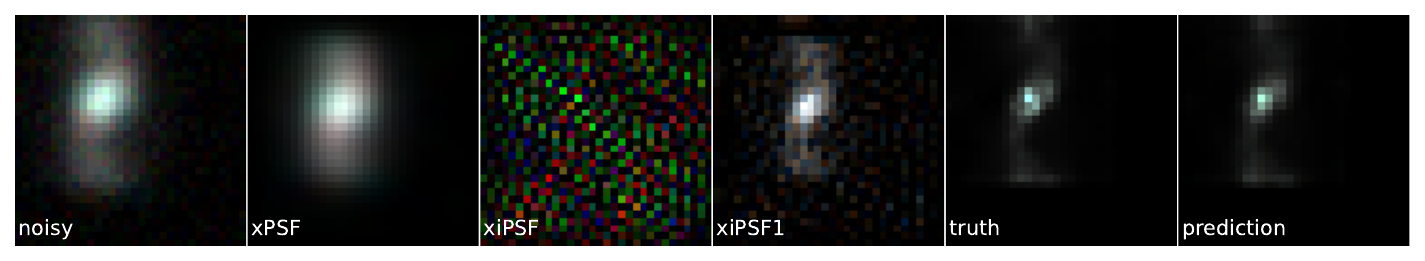}\\
\end{tabular}
} \\
\caption*{High SNR}
\end{minipage}\par\bigskip
    \caption{Figure showing some typical results from the \xPSF-\xiPSF-\xiPSFO \ network. The columns correspond to 1) noisy image, 2) noisy image convolved with the image PSF (\xPSF), 3) noisy image convolved with the inverse of the image PSF (\xiPSF) 4)noisy image convolved with the image PSF to which a Gaussian filter of standard deviation $1$ is applied (\xiPSFO), 5) truth image and 6) the prediction from the network. The rows are arranged according to the SNR as indicated. }
    \label{fig:ims}
\end{figure*}

\begin{table*}[htbp!]
\setlength{\tabcolsep}{4pt}
\centering
\scalebox{0.9}{
\begin{tabular}{l|cc|cc|cc|cc|cc|cc|cc cr}
  & \multicolumn{2}{c|}{\xPSF-\xiPSF-\xiPSFO} & \multicolumn{2}{c|}{\xPSF-\xiPSF} & \multicolumn{2}{c|}{\xiPSF-\xiPSFO} & \multicolumn{2}{c|}{\xPSF-\xiPSFO}& \multicolumn{2}{c|}{\xPSF}& \multicolumn{2}{c|}{\xiPSF}& \multicolumn{2}{c}{\xiPSFO}\\ 
\midrule
  & Mean   & Median & Mean  & Median & Mean &  Median & Mean &  Median & Mean & Median& Mean & Median & Mean & Median   \\ 
\midrule
PSNR & 75.177  & 74.744
& 73.477& 72.873
&74.056 &73.611
& 75.378&74.987
&74.034 &73.709
&73.331&72.842
&75.290&74.943
\\ 
1-SSIM & 0.0489  &  0.0177
& 0.0557& 0.0253
& 0.0507&0.0209 
& 0.0404&0.0165
&0.0628 &0.0259
&0.0708&0.0281
&0.0418&0.0173
\\  
\bottomrule 
\end{tabular}
}
\caption{Table showing the differences between the mean and median in the PSNR (dB) and SSIM values for the truth and predicted data sets for the different data cubes. For both PSNR \& SSIM, the higher the value, the better the congruence between the truth and the prediction.}
\label{tab:met}
\end{table*}

\subsection{(Re-)Normalization} \label{sec:renorm}

A perennial problem in the application of neural networks to astronomical image analysis is the image dynamic range. Dynamic range in astronomical images can vary by many orders of magnitude, while the neural networks require inputs to be of limited dynamic range. The non-linear transformations inside the neural network are typically non-linear over the range $0-1$  and if given an input range that is very different, the system will fundamentally change its properties. The usual procedure is therefore to apply certain normalization factors to the inputs, outputs as well as the truth images. The main issue is that one is ``not allowed'' to derive these factors from truth images, since for real-life problems we do not have truth images. Instead we argue that the normalization factor should be recalculated with an afterburner maximum likelihood for the amplitude parameter only.

In our case the procedure is as follows:
\begin{enumerate}
    \item Train the network with input and output images normalized in any sensible manner, e.g. using min-max rescaling to unity interval.
    \item When applying to the data, use the same rescaling on the input image as in the training. The output image is now deconvolved, but normalized in an arbitrary manner.
    \item Perform a maximum-likelihood fit to the output image normalization. In our case, this involves reconvolving the output image with the PSF and then calculating the renormalization factor as 
    \begin{equation}
        r = \frac{\sum_{({\rm pixels}\ i)}{n_i p_i}}{\sum_{({\rm pixels}\ i)}{p_i^2}}, 
        \label{eq:renorm}
    \end{equation}    
    where $p_i$ is the output image convolved with PSF and $n$ is the noisy image. This formula can be derived by minimizing the square difference between the $p$ and $n$.
\end{enumerate}

This approach leverages the neural network to do the actual heavy lifting in terms of determining the morphology of the output image, while leveraging the exact solution for the image normalization part. A similar approach can also be used in various deblending approaches employing neural networks, where perhaps multiple component amplitudes can be computed.

Upon implementing this procedure, we have found that the measured fluxes have small negative biases. This is due to the fact that any imperfection in the recovered shape will result in the object amplitude being biased low. To see this, imagine if the image to be fit in amplitude has a source at a completely wrong position. That source would not be able to model the flux of the actual source and therefore its flux would be low (and presumably consistent with the upper limit for the flux of a putative source at the wrong position). In our case, the situation is less dramatic, but sources are still biased low at a few percent. In order to fix this, we model the true total flux $M_{00}$ as a quadratic function in the recovered flux. The motivation for the quadratic function is that we expect morphology to be better recovered with higher fluxes and therefore less biased.

\section{Experiments \& Results} \label{sec:e&r}

\subsection{Deconvolution performance} \label{subsec:dp}
We start by conducting experiments with different combinations of the aforementioned \xPSF, \xiPSF \ and \xiPSFO  \ data configurations. Some typical results using the data cube \xPSF-\xiPSF-\xiPSFO \ are represented in Figure \ref{fig:ims}. As indicated in the postage stamps, the columns are noisy data, \xPSF, \xiPSF, \xiPSFO, truth and prediction respectively, arranged by the image SNR. Please note that in this work we use $M_{00}$ as a proxy for SNR, we have observed that they are analogous as far as flux recovery is concerned. Moreover, in terms of this work, $M_{00}$ is a more tangible quantity and it makes more physical sense to use it. In order to observe the behaviour in different SNR regimes, we split our data set based on the $M_{00}$ values of the truth data set. We have divided the data set into three in such a way that each bin contains approximately the same number of objects (3334, 3333 \& 3333 respectively). In terms of the values, the boundaries of the bins are $1.45 - 121.55$ (low), $121.55 - 562.22$ (medium) and $562.22 - 3739.12$ (high) respectively.

We would like to note that these are only the results for the non-blank objects, obtained from a network that was trained with blank images as well. 

We see that the network is capable of successfully deconvolving fine structures well below the size of the PSF. The \xiPSFO \ image is close to a formally regularized solution (e.g. Wiener-filter like). For low SNR objects in Fig \ref{fig:ims} it under-regularizes, as indicated by still significant noise features, while for the high SNR it over-regularizes. In any case, the prediction can correctly deconvolve fine structures in the truth image even in the low-noise regime. These images show that as the basic level, the network performs as one would expect. We next turn to more quantitative assessments of its performance.

In Table \ref{tab:met}, we compare the mean and median values of certain standard metrics that are typically used in image analysis for all the data combinations. Peak signal-to-noise ratio (PSNR) is defined as the ratio of maximum possible signal strength to the distorting noise level, expressed in logarithmic decibel scale. It performs a pixel-to-pixel comparison between images, and higher values of this quantity are equated to better quality, and lower values to greater numerical dissimilarities between images (\cite{HZ2010}). SSIM on the other hand, defines the perceived similarity between images through the correlation of image pixels, thus providing information about contrast, luminance and structure. It maps the structural similarities between two objects, and is often taken as an acceptable proxy to human visual perception, and is as such, considered to be more sensitive to image degradation as a result of compression, noise etc. than PSNR. Here we have used the implementations in \texttt{scikit-image} (\cite{van2014}) to calculate both PSNR and SSIM. In general, the higher the PSNR and SSIM values are, the better. Here, for SSIM we have opted to show how dissimilar the truth and prediction images are (1-SSIM since the maximum possible value is 1) rather than the values themselves, so the lower the value the better the similarity between the truth and prediction. For both PSNR and SSIM, the best mean and median values is returned by \xPSF-\xiPSFO, closely followed by the full \xPSF-\xiPSF-\xiPSFO \ while \xiPSF \ returns the worst mean and median values.

\begin{figure}
    \includegraphics[width=0.4\textwidth]{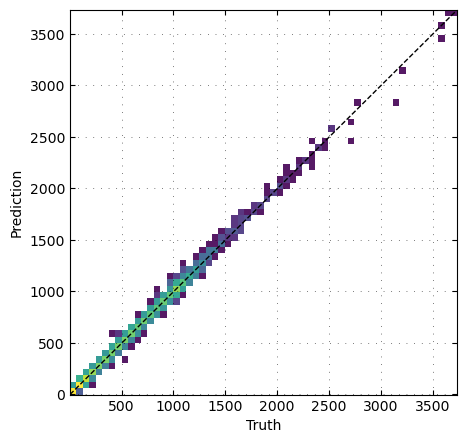}
    \caption{2-D histogram showing the $M_{00}$ moments of the raw truth data (X-axis) and the re-scaled and flux corrected predictions (Y-axis). We use this to illustrate the accuracy of our re-scaling method and flux corrections.}
    \label{fig:m00}
\end{figure}

\begin{table*}[htbp]
\setlength{\tabcolsep}{4pt}
\centering
\scalebox{0.9}{
\begin{tabular}{l|cc|cc|cc|cc|cc|cc|cr}

  & \multicolumn{2}{c|}{\xPSF-\xiPSF-\xiPSFO} & \multicolumn{2}{c|}{\xPSF-\xiPSF} & \multicolumn{2}{c|}{\xiPSF-\xiPSFO} & \multicolumn{2}{c|}{\xPSF-\xiPSFO} & \multicolumn{2}{c|}{\xPSF}& \multicolumn{2}{c|}{\xiPSF}& \multicolumn{2}{c}{\xiPSFO}\\
\midrule
  & Mean   & RMS & Mean  & RMS & Mean &  RMS & Mean &  RMS & Mean & RMS &Mean & RMS & Mean & RMS \\
\midrule
$M_{00}$ & 0.4313 & 17.6943
&  0.8253 & 24.4059
& 0.6633 & 24.5903
& 0.9567 & 17.9097
&-0.5118 & 26.4617
&0.1937 & 26.2412
&0.9077& 18.5814
\\
$\left<x\right>$ & 0.0044 &  0.6096
&0.0293 &0.6314 
&0.0201&0.6543 
&0.0392 &0.5610
&0.0237 &0.5391
&0.0192&0.8376
&0.0337&0.6029
\\
$\left<y\right>$ & 0.0070 & 0.6034
& -0.0166 & 0.6204
&0.0033 &0.6282
&-0.0070 &0.5436
&-0.0229 &0.5486
&0.0365&0.863
&0.0144&0.5575
\\
$e_{1}$ &-0.0005& 0.1443
&-0.0050 & 0.1503
&-0.0082&0.1424 
& -0.0030&0.1376
& 0.0108&0.1541
&-0.0043&0.1585
&0.0010&0.1350
 \\
$e_{2}$ &0.0042&0.1586
&-0.0147 &0.1719
&0.0040 &0.1618
& -0.0011&0.1561
& -0.0117&0.1717
&-0.1120&0.1862
&-0.0114&0.1605
\\
$\|\boldsymbol{e}\|$&0.0670&0.1425
& 0.0581& 0.1465
&0.0486 &0.1356
& 0.0444&0.1308
&0.0908 &0.1698
&0.0729&0.1581
&0.0425&0.1281
\\
$s$& -4.3078&11.9458 
&-2.8706 & 12.2596
& -2.4115&12.2552
&-0.3480&9.2682
& -8.7653&16.4317
&-6.7513&18.1492
&-0.8512&9.1547
\\
\bottomrule
\end{tabular}
}
\caption{Table showing the mean and RMS values of the differences in moments \& other astronomically significant quantities between the true and the predicted images for the different data combinations. }
\label{tab:mom}
\end{table*}

Table \ref{tab:mom} shows a comparison of quantities that are physically significant in astronomical image processing in terms of image moments (see Equation 10 in \texttt{HW23}). We have tried to quantify our results in terms of the four major image aspects, i.e., (1) position, (2) flux, (3) shape and (4) size. Here, the central moment $M_{00}$ is a measure of the total flux, $<x>$ \& $<y>$ astrometric co-ordinates ($M_{10}$ and $M_{01}$ moments), the second central moments $x^2$, \& $y^2$ ($\mu_{20}$ and $\mu_{02}$) representing size, $s = x^2 + y^2$, and $e_1$ \& $e_2$ the $+$ and $-$ components of ellipticity, and $\|\boldsymbol{e}\| = e_1^2 + e_2^2$. Since these are the differences between the mean and RMS values of the truth and prediction, the objective is that the absolute value of the difference be as small as possible.

We apply corrections to the recovered $M_{00}$ values as described in the Section \ref{sec:renorm}. The resulting function for the \xPSF-\xiPSF-\xiPSFO \ has linear and quadratic coefficients equal to $0.984$ and $2.84 \times 10^{-6}$ and a constant term $-1.6$. The error percentage or bias between the predictions and this corrected data are $7.25\%$, $2.15\%$ and $1.5\%$ respectively for the low, medium and high SNR bins and those between the truth data and the corrected predictions are $1.46\%$, $0.22\%$ and $0.066\%$, well within the noise scatter and consistent with no bias. 

In Figure \ref{fig:m00}, we illustrate these results graphically, we plot a 2-d histogram of the $M_{00}$ values for the raw truth data and the prediction values rescaled using the factors calculated in Equation \ref{eq:renorm}  and then flux corrected for the \xPSF-\xiPSF-\xiPSFO \ data cube. As can be seen,  the $M_{00}$s are within the same range and to a great extent align with each other. We single out the $M_{00}$ moment here to illustrate this effect since all the other quantities that we consider are flux scale independent. Even though only one of the metrics we consider is affected by it, the impact of re-scaling cannot be ignored because $M_{00}$ is representative of the total flux and is therefore quite important in terms of image recovery. We note that all fluxes are biased low. That is because any shape mismatch between the output image and the truth will result in normalization being low.

The best values for both the mean $M_{00}$ (denoting bias) is returned by the \xiPSFO \ data cube, and the RMS (scatter) by \xPSF-\xiPSF-\xiPSFO, the worst mean by \xPSF-\xiPSFO, and the worst RMS by \xPSF \ with the other data configurations performing moderately. The best mean value for $\left<x\right>$ is returned by \xPSF-\xiPSF-\xiPSFO \ and that for $\left<y\right>$ by \xiPSF-\xiPSFO. The latter is quite interesting because convolving noisy image with the image PSF (\xPSF) is a matched filter for point sources and therefore often used for object detection in astronomical images, but here the object detection aspect seems to be achieved better (albeit marginally) without the \xPSF \ component. It is also worth noting that for mean $\left<y\right>$, \xPSF-\xiPSF-\xiPSFO \ and \xPSF-\xiPSFO \ have remarkably similar values while for $\left<x\right>$ the values of \xPSF-\xiPSF-\xiPSFO \ is an order of magnitude better than all others. For the RMS value of the astrometric co-ordinates however, \xPSF \ returns the best (lowest) values. All other data cubes exhibit similar performances except for \xiPSF \ which performs the worst.

 The best mean value for the $+$ component of ellipticity, $e_1$, is returned by \xPSF-\xiPSF-\xiPSFO \ and the worst by \xPSF; likewise the best RMS value is returned by \xiPSF1, closely followed by \xPSF-\xiPSFO. For the $-$ component of ellipticity, $e_2$, the best mean value is returned by \xiPSF-\xiPSFO \ \& \xPSF-\xiPSF-\xiPSFO \ (the difference between them only 0.0002) and the worst by xiPSF. The latter is a similar result to that of the \xPSF \ being inept at determining object position, because convolution with an inverse PSF (\xiPSF) is generally implemented for shape recovery in images. The best RMS value for $e_2$ was also returned by \xiPSF-\xiPSFO \ and the worst by both \xPSF-\xiPSF \ \&  \xPSF \ (also 0.0002 difference). For the aggregated ellipticity, $|\boldsymbol{e}\|$, the best mean and RMS values are returned by \xiPSFO \ and the worst of both by \xPSF. For the quantity $s$, denoting the sizes, the best mean value is returned by \xPSF-\xiPSF, the best RMS by \xiPSFO, the worst mean by \xPSF \ and the worst RMS by \xiPSF.

\begin{figure*}[htbp]
\centering
\includegraphics[width=\textwidth]{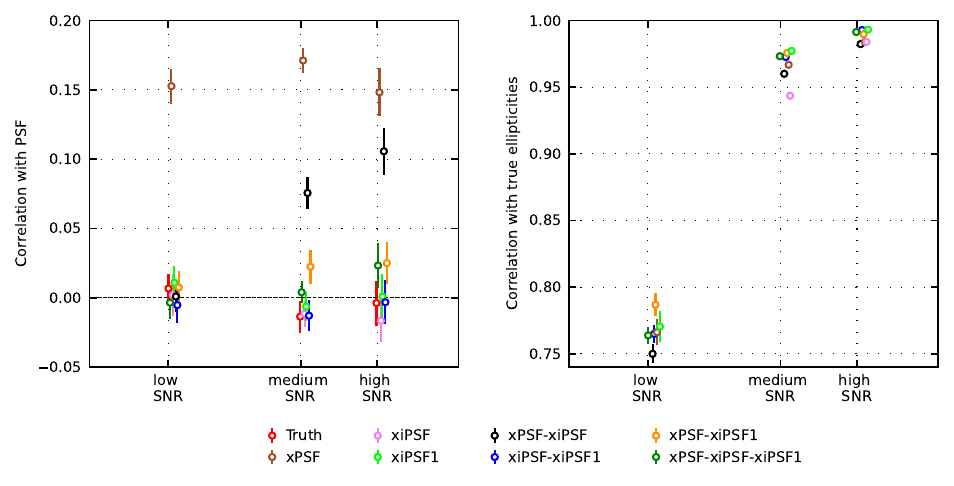}
\caption{Plot showing the residual PSF dependence (left) and correlation with the truth data (right), separated by SNR. The different data combinations are indicated in the legend below, and they are slightly offset in the X-axis for better visualization. }
\label{fig:corrs}
\end{figure*}

In Figure \ref{fig:corrs}, we explore the output ellipticity correlations with PSF \& truth image  using Equations (29) and (30) in \texttt{HW23} (Eqs. 29 \& 30 hereafter). The left panel denotes the PSF correlation of the predicted and the truth data sets for the different experiments that we have conducted.  For an ideal deconvolution, this correlation should be zero. 

This is an important quantity since in applications like shear estimation, presence of residual PSF dependence could lead to `additive bias' where shapes are contaminated by PSF (see \cite{RM2018} and references therein). An ideal network should remove all PSF dependence in the predicted image, leaving the quantity defined in Eq. 29, $\xi_{PSF} = 0$. In order to verify this, we also examined the recovered shapes of objects in relation with the truth image as defined in Eq. 30. For an ideal case this quantity, $\xi_{true} = 1$ and this is represented in the right panel of the figure. The different data configurations used as input to the network are denoted in the legend. As is evident from the left panel of the figure, it seems that just \xPSF \ and \xPSF-\xiPSF \ both do not seem to provide sufficient  and information for the network to remove the residual PSF dependence. All the other combinations seem to be comparable in this respect. We expect the \xPSF \ case to be the worst since the information is simply not present and we indeed find this to be the case.   The \xPSF-\xiPSF \ combination also fares quite badly, likely by being dominated by the information present in the \xPSF.  All the data combinations have similar performance with regard to ellipticity recovery as can be seen from the right panel of Figure \ref{fig:corrs} with marginally better performances by \xPSF-\xiPSFO \ in the low SNR bin, and \xiPSFO \ in the medium and high SNR bins.  For convenience of visualization, the points are slightly offset on the X-axis. 

\subsection{Object detection efficiency}

\begin{figure*}[htbp]
    \centering
    \subfloat[\xPSF]{\includegraphics[width=0.5\textwidth]{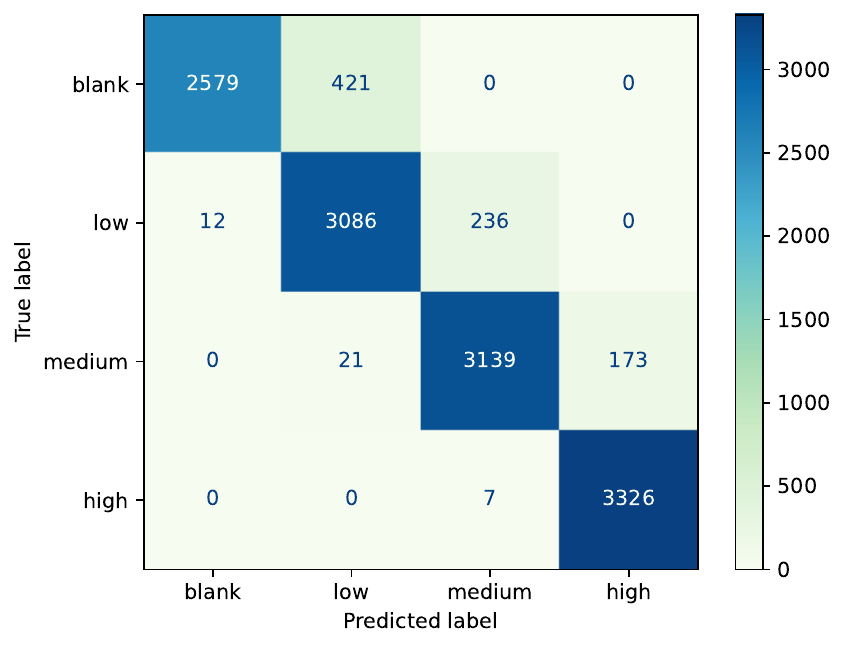}}
    \subfloat[\xiPSFO]{\includegraphics[width=0.5\textwidth]{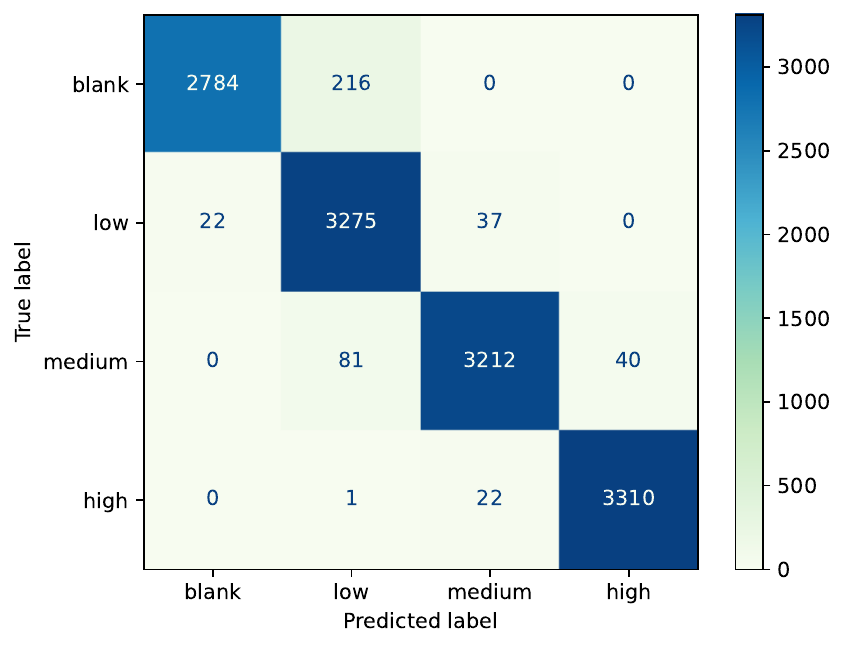}} \\
    \subfloat[\xPSF-\xiPSF]{\includegraphics[width=0.5\textwidth]{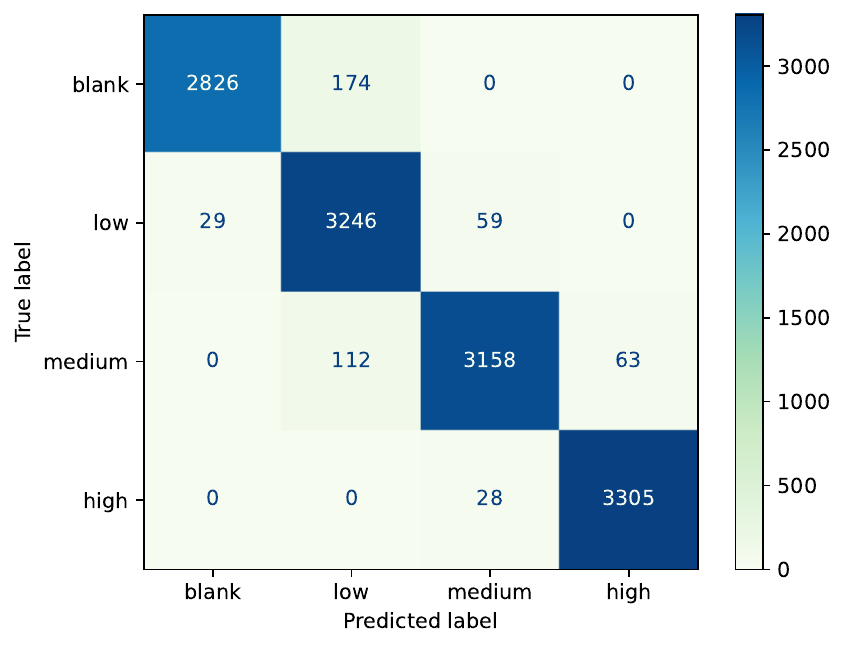}}
    \subfloat[\xPSF-\xiPSF-\xiPSFO]{\includegraphics[width=0.5\textwidth]{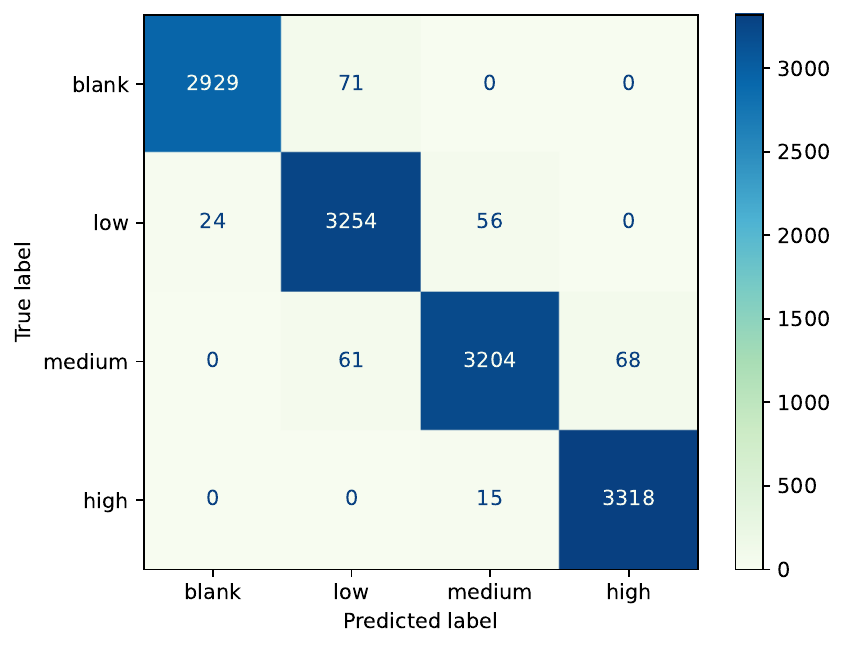}}\\
          \caption{Figure showing the confusion matrices for a few selected networks. The true labels are on the Y-axis and the predicted labels on the X-axis. }
\label{fig:cm1}
\end{figure*}

Figure \ref{fig:cm1} shows the confusion matrices for a few selected networks, starting with \xPSF \  (our worst performing network) to the complete \xPSF-\xiPSF-\xiPSFO \ network. Confusion matrices are a standard representation method  for the results of machine learning algorithms, especially classification problems. In a typical $2 \times 2$ confusion matrix used for binary classification problems, the values represented are the true positive rate (TPR; predictions and truth correspond), false positive rate (FPR; predictions classify as positive while the truth disagrees) from left to right in the first row; the false negative rate (FNR; predictions classify as negative while the truth disagrees) and the true negative rate (TNR; predictions and truth agree that it's negative) from left to right in the second row. The left diagonal represents the correct classifications. In our case, we can divide our data set into four classes -- blanks, low SNR, medium SNR and high SNR. In reconstruction problems such as ours, a threshold needs to be defined to classify the predictions into detections or non-detections. We simply defined this threshold as the boundary values for the low, medium and high SNR bins (mentioned in Section \ref{sec:e&r}). 

Part (a) of Figure \ref{fig:cm1} represents the confusion matrix for the network trained solely using \xPSF. For the high SNR bin, this network seems to return the best results, which makes sense intuitively, since noisy images are often convolved with their respective PSFs for object detection, and when the contrast between the signal and the noise is higher, this must make the task easier. This reasoning is supported by the diagonal of the confusion matrix where it is evident that the detection efficiency increases with the increase in SNR. The blank vs low confusion seems to be prevalent in the case of all the different networks with varying degrees, it seems especially worse in the case of \xPSF \ with a blank detection accuracy of $\sim 86\%$. The network trained with \xPSF-\xiPSF-\xiPSFO \ (part (d)) seems to perform the best in this respect with an accuracy of $\sim  98\%$. There seems to be more tension between the true low and the predicted medium bins in the \xPSF \ than in any other network. The best performance in both the low and medium bins is returned by \xiPSFO (part (b)), with accuracies of $\sim 98\%$ and $\sim 96\%$ respectively. Except for the \xPSF \ network, all the others seem to have a slight issue when dealing with the medium bin, as is evidenced by the lower accuracy compared to the other bins (excluding the blanks). There doesn't seem to be any trend as to how this issue manifests, whether it is an overestimation or underestimation. Part (d) of the figure shows the results from the network trained on \xPSF-\xiPSF-\xiPSFO, and it returns a substantially higher accuracy compared to the other networks in the blank bin ($\sim 98\%$). All the networks perform very well in the high SNR bin, all with $\sim 99\%$, which is not a surprise since the aforementioned contrast between signal and noise must be the most apparent in the case of these objects.

From both Figure \ref{fig:cm1}  and from Tables \ref{tab:met}--\ref{tab:mom} we can comfortably conclude that the inclusion of \xiPSFO, while unconventional, is definitely helpful in both detection and shape recovery, especially in the low SNR regime. It is also similarly evident that only \xPSF \ or only \xiPSF \ simply does not contain sufficient information for this purpose. As can be seen from Figure \ref{fig:ims}, even visually it appears that \xiPSFO \ encompasses salient features of both \xPSF \ and \xiPSF, which could be the reason why it is more suited to the purpose at hand. It is also possible that this is data dependent -- while we have tried to make our data set as varied as possible, this method might perform differently in a distinct data set, requiring certain adjustments, for instance, the Gaussian regularising filter might have a different FWHM. The size of the data cube is not linearly related to the running time required, either way the network takes between 2 and 3 hours to complete, which provides a solid scope for experimentation with the data manipulation parameters. 

\begin{table*}[htbp]
\setlength{\tabcolsep}{4pt}
\centering
\scalebox{0.95}{
\begin{tabular}{l|cc|cc|cc|cc|cc|cc|cr}

  & \multicolumn{2}{c|}{\xPSF-\xiPSF-\xiPSFO} & \multicolumn{2}{c|}{\xPSF-\xiPSF} & \multicolumn{2}{c|}{\xiPSF-\xiPSFO} & \multicolumn{2}{c|}{\xPSF-\xiPSFO} & \multicolumn{2}{c|}{\xPSF}& \multicolumn{2}{c|}{\xiPSF}& \multicolumn{2}{c}{\xiPSFO}\\

\midrule
  & sources   & blanks & sources  & blanks &sources &  blanks & sources &  blanks & sources & blanks &sources & blanks & sources & blanks \\
\midrule
&10047 & 2953
&10145 & 2855
&10210 & 2790
&10193 & 2807
&10409 & 2591
&10163 & 2837
&10194 & 2806
\\

\\
\bottomrule
\end{tabular}
}
\caption{Table showing the distribution of detected sources and blanks with each data cube. Please note that these are the total number of objects the networks class as sources or blanks, the misclassification into the different SNR bins is not considered (e.g. if an object with low SNR is classed by the network as medium or high, it is still included as `source').}
\label{ratios}
\end{table*}

In Table \ref{ratios}, we show the statistics of source and blank detections for the different data cubes and corresponding networks. Please note that all the objects the networks identify as not blanks, even if they are an overestimation or underestimation of flux (e.g. an object of true class low classified as medium or vice versa), are classed here as `sources'. The general trend from the numbers seems to be a confusion of the blank images being reconstructed as source images, rather than vice versa. We confirmed this by examination of a subset of the truth and reconstructed images. We can only speculate as to the cause of this, it is possible that for a small subset of the images, the convolutions and the addition of noise could give rise to the appearance of flux which causes the overestimation by the network. Since the percentage of this type of misclassification seems quite small, it might be possible to rectify this simply by increasing the size of the data set. 

\section{Summary \& Discussion} \label{sec:summconcl}
We present in this work, a simpler approach to remove residual PSF dependence and recover ellipticity as a follow-up to the work done in \texttt{HW23} with a more data-driven approach. The network implemented in this work is a convolutional autoencoder with 3 encoder and correspondingly 3 decoder layers with respective filter sizes of 64, 128 and 256. $5 \times 5$ kernels are used in the convolutional layers (both encoder and decoder) while $2 \times 2$ kernels are used in the pooling layers in the encoder and upsampling layers in the decoder. All convolutional layers are activated with the \texttt{LeakyReLu} function and the final output layer with a \texttt{softplus} function.  The model is compiled with the \texttt{Adam} optimiser and the loss function used is \texttt{BinaryCrossentropy}. We trained this network with several data configurations, namely, \xPSF, \xiPSF, \xiPSFO \ and combinations thereof, results of which are detailed in the preceding sections. 

We quantify our results using various metrics, which are displayed in Figures \ref{fig:ims} - \ref{fig:cm1} and Tables \ref{tab:met} - \ref{tab:mom}. In Figure \ref{fig:ims}, we show some typical results that is output by the network that is trained with a \xPSF-\xiPSF-\xiPSFO \ data cube/combination as a proof of concept. We analyse this further with specific correlations between the predicted data and the PSF and between the predicted data and the true ellipticities in Figure \ref{fig:corrs} in the three SNR bins. In Tables \ref{tab:met} and \ref{tab:mom}, we compare significant quantities, both from image analysis and astronomical perspectives to quantify the similarities between the predictions and the truth data. In Table \ref{tab:met}, the quantities compared are PSNR and SSIM which are popular metrics used to evaluate image reconstruction. While PSNR and SSIM provide some insight into image recovery from an image analysis standpoint, we are more interested in the recovery of astronomically significant features. To this end, in Table \ref{tab:mom}, we compare the differences between moment proxies for total fluxes, first order and second order moments averaged over the three bands for the different data combinations.  Then we analysed our results using confusion matrices, a very popular form of representation generally used in classification problems by defining thresholds based on $M_{00}$ (used analogously to SNR), as shown in Figure \ref{fig:cm1}.

We conclude that as is general convention, just convolving the noisy image with the PSF (\xPSF) or with the inverse PSF (\xiPSF) by themselves are not informative enough for the autoencoder network to efficiently locate the object, remove the PSF dependence and recover the original shape. \xiPSFO \ on the other hand, provides very good results, both by itself and when used in combination with \xPSF \  or \xiPSFO \ (or both). 

It is hard to declare a clear winner. \xiPSFO \ is the most economical and is close to the best, although on certain tests, augmenting it with the other two combinations improves results. The simplest interpretation is that \xiPSFO \ provides a blurry image that is standardized across various observed PSFs -- the role of the network is then to simply sharpen and denoise it based on how galaxies in the training set look like. As in any denoising/deblurring approach, the missing information is filled-in based on priors and therefore one should be cautious not to over-interpret galaxy morphologies derived this way.

We did consider using the noisy images directly as input to the network(s), just in case it contained extra information that might be useful for both flux and shape recovery, but found that the impact was not favorable. Table \ref{incblur} shows the results for our metrics when the blurred image is also used as input alongside \xPSF \ and \xiPSF \ (as is mentioned earlier, \xPSF \ and \xiPSF \ are generally used for detection and shape recovery (unregulated deconvolution) respectively). As is evident from the values, providing the blurred/observed image seems detrimental in both flux and shape recovery (as indicated by the $M_{00}$ \& $s$ values). The noisy nature of these images could be making both object detection and shape recovery difficult. 

As with all kinds of neural network applications, the more data is given as input the better the performance would undoubtedly become. We ran experiments with the \xPSF-\xiPSF-\xiPSFO \ network with different number of objects as input, the results of which are given in Table \ref{diffsize}. Please note that the numbers given (100k, 10k etc.) denote the number of images with sources in them, to which a $30\%$ ratio of blank images are added. The $M_{00}$ values in the table show how the size of the training sample impacts the flux recovery – as can be seen, the flux recovery RMS values improve (gets lower) as the size of the training set increases. It can be inferred from the RMS values for all the metrics that we are getting closer to convergence as the size of the data set increases, and it is reasonable to assume that with a larger data set this value would decrease further.

We have refrained from comparisons of our results to our previous work, \texttt{HW23} because both methods are `philosophically different’, \texttt{HW23} having an algorithm-driven approach while this work has a data-driven approach. As we have mentioned in Section \ref{sec:data}, we have modified the dataset generation slightly in this work, and a direct comparison is not beyond the realm of possibility. This is something that we might explore in the future. We have also chosen not to offer comparisons with standard analysis methods such as SExtractor or PSFExtractor \citep{psfex} because of the technical differences between the approaches such as the absence of an element of deconvolution of the PSF which renders the comparison physically insignificant while conceding that there are several routes to the objective of removing PSF dependence and measuring object parameters from astronomical images. For instance, there are PSF modelling techniques that could possibly be used as a filter to remove PSF dependence in wide-field survey images, such as those generated by \citep{herbel2018} and the adaptive optics PSF estimation toolbox DEEPLOOP \citep{deeploop2022}. \cite{herbel2018} is especially relevant and interesting because of its forward modelling approach using a combination of Singular Value Decomposition (SVD) and a CNN to extract the features of the PSF using  r-band stellar images from Sloan Digital Sky Survey Data Release 14(SDSS DR14, \cite{sdssdr14}). There also exist, approaches such as in \citep{stoppa2023} which describes parameter estimation using a two-step variance estimation network without any knowledge of the PSF (with caveats). 

The main advantages that we see for this method that we have put forth are, (1) simple neural network (2) fast processing, (3) minimal data manipulation and (4) minimal reliance on data features that might not be readily available (e.g. normalization factors). We would also like to emphasize that this method is not a `be all and end all' product in removing PSF dependence and ellipticity recovery by any means. This is merely a simpler way to obtain fast results in this respect using marginal data manipulation. It is possible that its performance will vary in a different data set. While we haven't tested it on other data sets, we feel that it is not suitable for transfer learning purposes. This is because our network architecture is relatively simple (only 3 encoder-decoder layers) and trained on a relatively smaller data set (130,000 images) in contrast with models that are generally used in transfer learning problems such as U-Net \citep{unet} or ResNet \citep{resnet} both of which have complex architectures and are trained on a large amount of diverse data of million(s) of images. Although the transfer learning requirement might be rendered moot by the fact that this is a very fast method even on a data set that contains some 100,000 objects, so training on a similar sized data set would also be a fast process. However, it is an avenue that we might explore in future work. Our only claim in this work is that for similar data that has both noise and contains one centred object, this is an effective, albeit crude, specific method to obtain denoised data that has excellent correlation to the truth image. There are certain approaches that might be applied to refine this method further. For example, it might be interesting to replace the CAE with a VAE -- which would add both some probabilistic and Bayesian aspects to the denoising process. Moreover, VAEs  also provide more tuneable parameters that give more control over the latent representation of the input data. We deliberately didn't consider `blended' or `overlapping' sources in this work, since this is meant as more of a `proof of concept' of using convolutional autoencoders for PSF removal. It might work `out of the box', or it might need a larger data set and/or a more complex network, but it certainly is an intriguing prospect for future work.

\bibliography{ref.bib}

\appendix

\section{Choice of loss function}

\begin{table}[h]
\setlength{\tabcolsep}{4pt}
\centering
\scalebox{0.95}{
\begin{tabular}{l|cc|cc|}

& \multicolumn{2}{c|}{\xPSF-\xiPSF-\xiPSFO (BCE)} & \multicolumn{2}{c|}{\xPSF-\xiPSF-\xiPSF (MSE)} \\
\midrule
  & Mean   & RMS & Mean  & RMS \\
\midrule
$M_{00}$& 0.4313 & 17.6943
&-1.4905&65.9536
\\
$\left<x\right>$ & 0.0044 &  0.6096
& 0.0802 & 1.4667

\\
$\left<y\right>$ & 0.0070 & 0.6034
& 0.2531 & 1.5209
\\
$e_{1}$ &-0.0005& 0.1443
&0.019 & 0.2179
\\
$e_{2}$ &0.0042&0.1586
&-0.1695 & 0.3353

\\
$\|\boldsymbol{e}\|$&0.0670&0.1425
&0.1877 & .2588
\\
$s$& -4.3078&11.9458 
&-33.1264 & 49.3645

\\
\bottomrule
\end{tabular}
}
\caption{Table showing the comparison of the metrics with different loss functions. BCE is the loss function used across this work, MSE is a popularly used loss function in neural networks. There is no general rule for choosing one or the other, but in our case, with our data set and network architecture, BCE is the one that works better.}
\label{bcemse}
\end{table}

Loss functions are a critical part of neural network based machine learning methods because they often determine the direction the training takes, by decreasing the disparities between the actual and predicted values. The function that we have used in this work, BCE, is generally used for image classification rather than image reconstruction problems. However, the far reaching consensus seems to be that there is no straight answer when choosing loss functions for autoencoders. Therefore we experimented with combinations of different loss functions and optimizers to find the one that worked best for our data set and network combinations, which was BCE with Adam optimizer on a learning rate schedule (0.0001 falling exponentially at the rate of 0.6 per 100,000 steps). An example of one of the experiments is given in Table \ref{bcemse}, showing the changes in our metrics when the loss functions are BCE and mean squared error (MSE) respectively for the same data, network architecture and optimizer. As can be seen, the values seem to be much better when the loss function is BCE. 

\begin{figure*}[htbp!]
    \centering
    \includegraphics[width=0.9\linewidth]{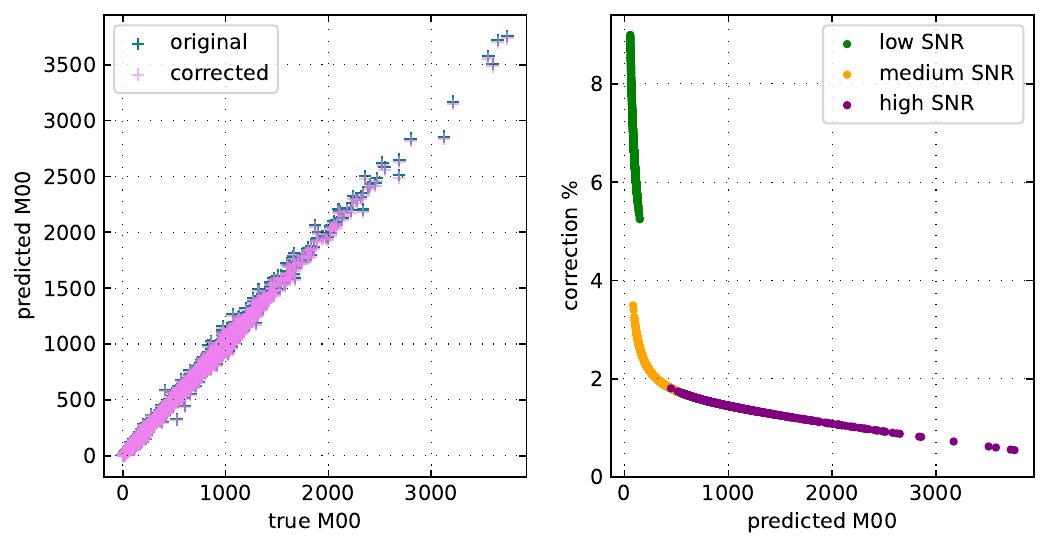}
    \caption{Plots showing the scaling of the predicted $M_{00}$ values with the true $M_{00}$ (left) and the corrections to the predicted $M_{00}$ values (right). In the plot on the left, the original predicted values are plotted in teal and the corrected values are plotted in violet. The scatter between the true and predicted values decrease after the quadratic corrections are applied. In the right plot, the applied corrections in percentages are plotted against the predicted $M_{00}$ values separated by the different flux bins (low = green, medium = orange, high = purple). The mean corrections for the three bins are $7.25\%$, $2.15\%$ and $1.5\%$ respectively.}
    \label{fig:m00corrm00}
\end{figure*}

In a mathematical sense, when MSE is used as the loss function, it compares the numbers directly, whereas BCE defines logarithmic probability levels to determine whether an object falls into true or false categories. This makes BCE more sensitive to the predictions closest to maxima and minima wherein either the negative or the positive parts of the function are activated(for instance, when a pixel is far away from the actual source, the values get set to 0). However, when you have a number that is in between, both parts are partially activated. Because of this greater sensitivity, the flux is often better estimated. In short, the losses returned by BCE are higher than that for MSE, so in cases where you want to penalize the errors more, BCE is preferred. It has also been suggested that if there is an element of nonlinearity in the output layer (here with the \texttt{softplus} activation function in the output layer) then perhaps BCE is better suited since it simulates a multi label problem and tries to maximize the likelihood of the output emulating the input.

\section{Biases in flux recovery}

\begin{figure}[htbp!]
    \includegraphics[width=0.4\textwidth]{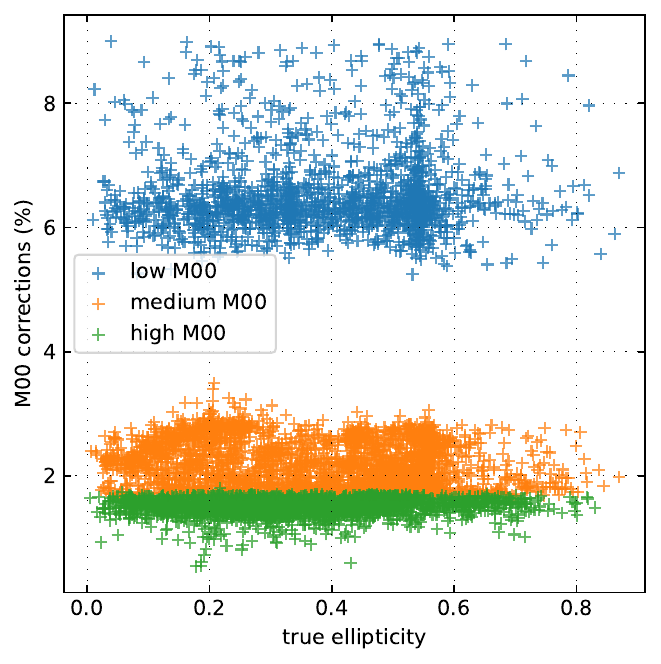}
    \caption{Plot showing the distribution of the corrections in percentages of the predicted $M_{00}$ values (Y-axis) vs the true ellipticities (X-axis). The corrections for the low, medium and high SNR bins are respectively represented in blue, orange and green. As is evident in the plot, there's no ellipticity dependence for the flux corrections, but there are distinct distributions for the low, medium and high flux bins as is expected. }
    \label{fig:evm00}
\end{figure}

As mentioned in Section \ref{sec:renorm}, we observe that the recovered fluxes are biased low in our predicted/reconstructed images, which we correct using a quadratic function with the coefficients mentioned in Section \ref{sec:e&r}. Figure \ref{fig:m00corrm00} shows graphical representations of this procedure. The left plot shows the predicted $M_{00}$s (teal) and the corrected $M_{00}$s (violet) plotted against the true $M_{00}$s. It can be seen that the scatter in the predictions is considerably reduced in the corrected predictions in comparison, indicating that they follow the true values more closely. In the right plot, we show the correction factors in percentages as derived from the quadratic function, separated by the three flux bins, plotted against the predicted $M_{00}$ values. As is mentioned in Section \ref{subsec:dp}, the mean of the correction percentages in the bins are respectively $7.25\%$, $2.15\%$ and $1.5\%$. The correction percentages are higher in the low SNR bin and becomes lower as the SNR increases. This stands to reason, it is both expected and observed that the network has an easier time reconstructing object images as the SNR increases. 
\begin{table*}[htbp!]
\setlength{\tabcolsep}{4pt}
\centering
\scalebox{0.95}{
\begin{tabular}{l|cc|cc|cc|cc|cc|}

& \multicolumn{2}{c|}{\xPSF-\xiPSF-\xiPSFO (100k)} & \multicolumn{2}{c|}{10k} & \multicolumn{2}{c|}{25k} & \multicolumn{2}{c|}{50k} & \multicolumn{2}{c|}{75k}\\
\midrule
  & Mean   & RMS & Mean  & RMS & Mean &  RMS & Mean &  RMS & Mean & RMS \\
\midrule
$M_{00}$ & 0.4313 & 17.6943
&  -0.721 & 43.8565
& 0.6002 & 27.5998
& 0.3629 & 20.0815
& 0.3321 & 18.3812

\\
$\left<x\right>$ & 0.0044 &  0.6096
& 0.2926 & 1.2453
&0.0129& 0.8429
&0.0102 & 0.6909
&-0.0231 & 0.6404

\\
$\left<y\right>$ & 0.0070 & 0.6034
& 0.2997 & 1.2445
&-0.05 &0.8293
&-0.007 & 0.6819
&0.0131 &0.6447

\\
$e_{1}$ &-0.0005& 0.1443
&- 0.007 & 0.187
&-0.0261 &0.1581
& -0.0011 &0.1465
& -0.0014 & 0.1441

 \\
$e_{2}$ &0.0042&0.1586
&-0.0878 &0.2574
&-0.0376 &0.2011
&-0.018 & 0.1794
& -0.0157 &0.1697

\\
$\|\boldsymbol{e}\|$&0.0670&0.1425
&0.1348 &0.2159
&0.0639 &0.1556
& 0.0559& 0.1425
& 0.0523 &0.1385

\\
$s$& -4.3078&11.9458 
&-12.6793& 27.6838
& -1.4411 &12.8687
& -3.4463 & 11.4015
& -3.4128& 11.3499

\\
\bottomrule
\end{tabular}
}
\caption{Table showing the values for our chosen metrics with data sets of different sizes for the same network architecture for the \xPSF-\xiPSF-\xiPSFO \ data cube. As can be seen, the flux recovery RMS values improve (gets lower) as the size of the training set increases. It is reasonable to assume that with a larger data set the detection and recovery will get even better.}
\label{diffsize}
\end{table*}

We also examined the shapes of the reconstructed images to see if they influence these correction factors, the result of which is represented in Figure \ref{fig:evm00}. Here, the correction factors are plotted against the ellipticities of the truth test images as an indicator of shape. It doesn't seem like there is a direct correlation between the corrections and the object shapes. 

Another factor that we considered was the size of the training set. The network was run with different subsets of the training set sizes with 10000 (+3000 blanks), 25000 (+7500 blanks), 50000 (+15000 blanks) and 75000 (+22500 blanks). Table \ref{diffsize} shows the results of these experiments for the metrics used in Table \ref{tab:mom} for these subsets. It is clear from the results of the $M_{00}$ RMS values that the size of the training set does impact the flux recovery in general and therefore indirectly the inherent biases, as they are seen to decrease as the size of the training set increases.

An important point to note is that these flux bias corrections are not universal. If any factors at all such as the PSF, the filters etc. would change, then this bias correction fitting would need to be repeated with the new images.

\begin{table}[htbp!]
\setlength{\tabcolsep}{4pt}
\centering
\scalebox{0.95}{
\begin{tabular}{l|cc|cc|}

& \multicolumn{2}{c|}{\xPSF-\xiPSF-\xiPSFO } & \multicolumn{2}{c|}{\blur-\xPSF-\xiPSF} \\
\midrule
  & Mean   & RMS & Mean  & RMS \\
\midrule
$M_{00}$& 0.4313 & 17.6943
&1.0835& 39.8106
\\
$\left<x\right>$ & 0.0044 &  0.6096
& -0.0679 &2.0394
\\
$\left<y\right>$ & 0.0070 & 0.6034
& -0.0101 &1.9719
\\
$e_{1}$ &-0.0005& 0.1443
&-0.0105 & 0.1889
\\
$e_{2}$ &0.0042&0.1586
&-0.0375 & 0.2343
\\
$\|\boldsymbol{e}\|$&0.0670&0.1425
&0.0768 & 0.1767
\\
$s$& -4.3078&11.9458 
& -21.4706 & 45.2105

\\
\bottomrule
\end{tabular}
}
\caption{Table showing a comparison of the metrics when the noisy images are also included in the data cube. As is evident from the values, providing the blurred/observed image seems detrimental in both flux and shape recovery (as indicated by $M_{00}$ \& $s$ values). This could be due to the noisy nature of the blurred images making both object detection and shape recovery tricky. This shows that there is no extra information present in the noisy images that was not present in the data cube used in this work.}
\label{incblur}
\end{table}

\end{document}